# Enhancement of Spintronic Terahertz Emission via Annealing in Ferromagnetic Heterostructures

YANG GAO[1†], YANBIN HE[1†], XIAOJUN WU[1*], TIANXIAO NIE[2,4*], CHUN WANG[3,5], DEYIN KONG[1], CHANDAN PANDEY[2], BO WANG[3,5], LIANGGONG WEN[2,4], CUNJUN RUAN[1], JUNGANG MIAO[1], LI WANG[3], YUTONG LI[3,5], AND WEISHENG ZHAO[2,4]

[1] Beijing Key Laboratory for Microwave Sensing and Security Applications, School of Electronic and Information Engineering, Beihang University, Beijing, 100191, China
[2] Fert Beijing Institute, BDBC, and School of Microelectronics, Beihang University, Beijing 100191, China
[3] Beijing National Laboratory for Condensed Matter Physics, Institute of Physics, Chinese Academy of Sciences, Beijing 100190, China
[4] Beihang-Goertek Joint Microelectronics Institute, Qingdao Research Institute, Beihang University, Qingdao, 266000, China
[5] School of Physical Sciences, University of Chinese Academy of Sciences, Beijing 100049, China
*Corresponding author: xiaojunwu@buaa.edu.cn; nietianxiao@buaa.edu.cn;

†These authors contributed equally to this work.

We systematically investigate the influence of annealing effect on terahertz (THz) generation from CoFeB based magnetic nanofilms driven by femtosecond laser pulses. Three times enhancement of THz yields are achieved in W/CoFeB through annealing effect, and double boosting is obtained in Pt/CoFeB. The mechanism of annealing effect originates from the increase of hot electron mean free path induced by crystallization, which is experimentally corroborated by THz transmission measurement on time-domain spectroscopy. Comparison studies of the thickness dependent THz efficiency after annealing are also implemented, and we eventually conclude that annealing and thickness optimization are of importance for scaling up THz intensity. Our observations not only deepen understanding of the spintronic THz radiation mechanism but also provide normal platform for high speed spintronic opto-electronic devices.

http://dx.doi.org/10.1364/AO.99.099999

## 1. INTRODUCTION

Extensive applications of terahertz (THz) technology call for superior emitters [1]. Femtosecond laser excited THz sources has been playing a critical role in the last three decades, stimulating various THz radiators [2-5]. Among them, recently developed spintronic THz emitter are predicted as one of the best candidates for the next generation THz sources, having a virtue for low cost, broadband, robustness as well [6-21]. Additionally, THz emission spectroscopy can be harnessed to deduce the femtosecond spin dynamics6. Consequently, more efficient THz sources can be realized. However, this new source is still in its infancy, and its mechanism needs to be explored thoroughly. Fundamentally, femtosecond laser pulse interaction with ferromagnetic hetero-structure excites electrons into sub-bands above Fermi level [22]. Due to the different mobilities of spin-up and spin-down electrons, a spin polarized longitudinal current is generated. When reaching capping layers, this current is converted to an in-plane transverse charge current, because of inverse spin Hall effect [6,7], and then radiates THz waves. In this process, material types (especially for spin Hall angles of heavy metals) and geometry structures have significantly impact on emission performance [7], and they have attracted much attention. However, fabrication conditions may also influence THz radiation [11]. For instance, Y. Sasaki et al. demonstrated that annealing on Ta/CoFeB/MgO films did enhance the THz emission intensity [11]. They proposed two possible explanations. The first one is the increase in spin-Hall angle and reduced hot electron velocity relaxion length caused by Boron atom diffusion into the Ta layer and crystallization of CoFeB with the help of MgO layer [23]. The second one is the increase of the hot electrons mean free path [11]. We are curious whether annealing effect can be extended to W or Pt capped CoFeB systems which has been well proven to have highly efficient.

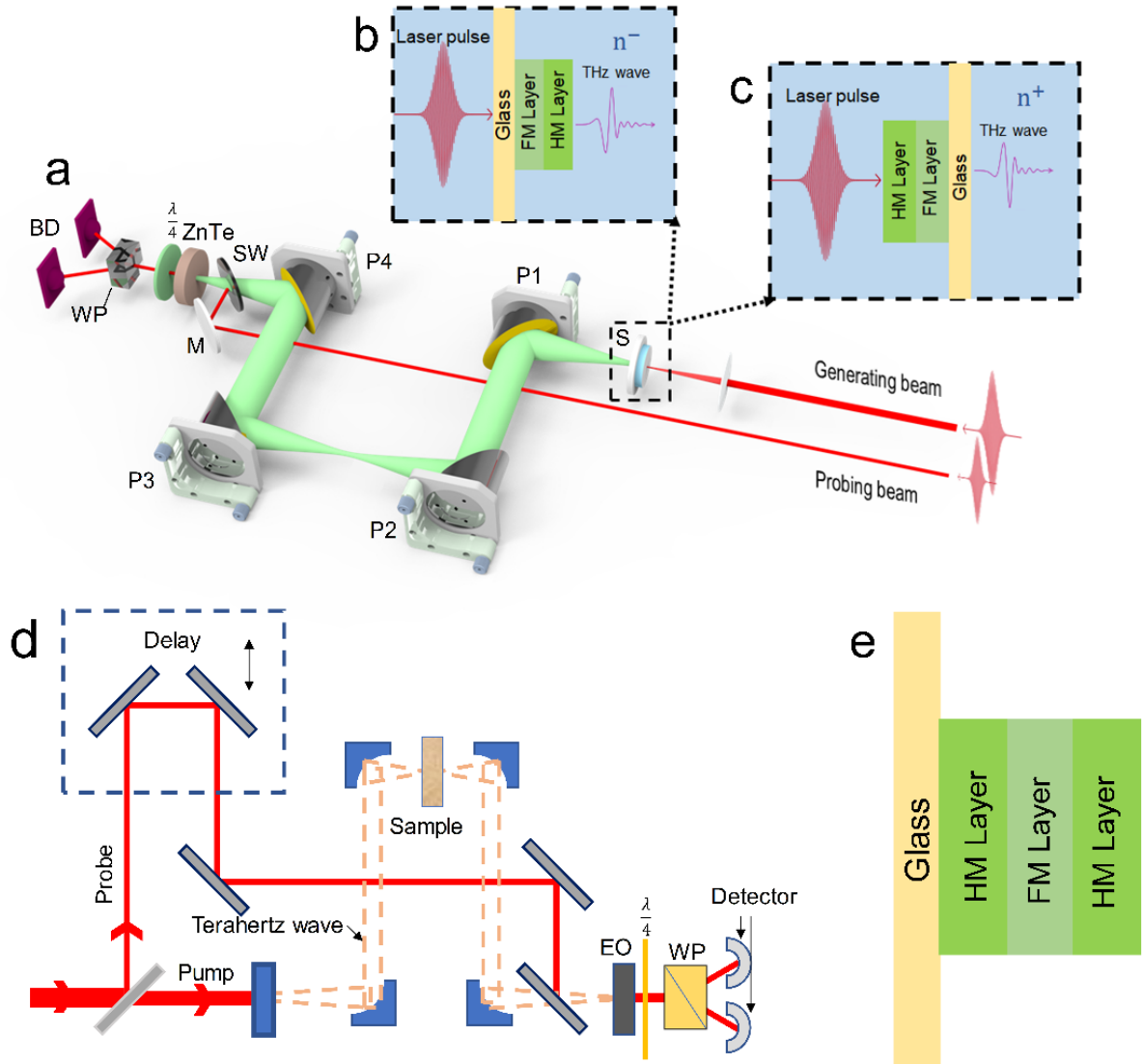

**Figure 1.** Schematic diagram of THz emission and transmission experiments and details of the sample structures. (a) Experimental setup of the THz emission system. BD: balanced detector; WP: Wollaston prism; λ/4: quarter wave plate; SW: silicon wafer for combining the probing beam together with the generated THz pulses; M: aluminum reflection mirror; P1-4: 90° off-axis parabolic mirrors; S: THz emitting samples. (b) and (c) definitions of $n^+$ and $n^-$. When the laser pulses first illuminate the glass substrate, it is defined as $n^-$, while the other case shown in Figure 1c is defined as $n^+$. Here, HM means heavy metal, including W and Pt. FM is ferromagnetic metal, which is CoFeB. (d) THz time-domain spectroscopy used in the transmission measurements. (e) Structure of the W/CoFeB/Pt trilayer sample.

Here, in order to verify the assumption, we conduct comparison emission experiments on W/CoFeB, Pt/CoFeB and W/CoFeB/Pt samples with and without annealing, which are different sample systems from those used in previous work [7]. We observe the magnitude of the THz waves generated from the materials after annealing is larger than that without annealing, and this enlargement can be reached up to 3 times. The following THz time-domain spectroscopy experiments indicate that transmitted THz waves in annealed materials does have a larger magnitude than the unannealed ones, which can in turn help understand the physical mechanism of annealing effect. Our results are helpful for quarrying and exploring the physical process of spintronic THz emission, optimizing the generation efficiency, furthermore, contributing to the high speed spintronic opto-electronic devices as well.

## 2. EXPERIMENTAL SETUP AND SAMPLE PREPARATION

Our experiment setups are schematically illustrated in Figure 1. Figure 1a is utilized in the THz emission experiments. The driven laser is a commercial Ti:sapphire laser oscillator yielding ultrafast pulses with the central wavelength of 800 nm, pulse duration of 70 fs, and repetition rate of 80 MHz. 90% of the laser power is used for generating THz waves while the rest for electro-optic sampling to record the generated THz temporal waveforms [24]. The generating beam is implemented to generate THz waves through the ferromagnetic heterostructure samples. The radiated THz signals are first collimated and then focused into a 1mm thick ZnTe detector. The THz electric field induces the variation of the refractive index in ZnTe which is indirectly recorded by the polarization rotation of the probing beam. Through combining a quarter waveplate, a Wollaston prism and a pair of balanced photodetectors, the generated THz temporal waveform is measured. Figure 1b and c are the enlarged arrangements of the femtosecond laser illumination onto the samples. The THz emitters as well as the THz optical paths are sealed in a vacuum to rule out the influence of water vapors. Figure 1d is the schematic diagram of the THz time-domain spectrometer, which is also driven by the same laser oscillator as the THz emission system. The THz pulses used to characterize the samples are generated from a photoconductive antenna and are probed by another 1mm thick ZnTe crystal. This system is purged by dry nitrogen gas.

We conducted the THz emission experiments on both CoFeB-based bilayers and trilayers which includes W(2.2)/CoFeB(2.0), W(4.0)/CoFeB(2.2), Pt(4.0)/CoFeB(2.2), W(2.0)/CoFeB(2.2)/Pt(2.0). The numbers in the bracket indicates the layer thickness in nanometers. These heterostructures were fabricated on 1 mm thick glass substrate in a high-vacuum AJA sputtering system with a base pressure of 10-9 Torr. The deposition conditions were carefully optimized to ensure the best quality and reproducibility and the growth rate for W, CoFeB, and Pt was 0.21 Å/s, 0.06 Å/s and 0.77 Å/s, respectively. We performed a sample rotation so as to ensure a good uniformity. In order to make comparison studies, all the samples are distributed into two groups, which are the as-deposited group and the as-annealed group. For the as-annealed group, samples were annealed at 280 °C in vacuum for one hour, and during this process, the magnitude of 1 T was set by a static external magnetic field. When conducting THz emission experiments, the applied static magnetic field is 50 mT. And for the THz transmission experiments, no external magnetic field is applied.

## 2. RESULTS AND DISSCUSSION

Figure 2a exhibit the THz signals from W(2.2)/CoFeB(2.0) nanofilms with and without annealing when the femtosecond laser pulses are incident on the positive side of the sample. The experiment results manifest that the annealing effect does improve the THz emission intensity. The ratio of the peak to peak values between the sample with and without annealing is $P+_{Annealing}/P=2.9$. This enhancement behavior is also observed when the pump laser pulses are incident on the negative side of $n^-$. In this case, as shown in Figure 2b, the THz intensity is as large as 3.2 times of the single from the unannealed sample. Furthermore, we also observe the THz emission polarity reversal behavior, which verify the directionality symmetry property of the inverse spin Hall effect mechanism. Likewise, the annealing effect induced THz emission enhancement is also detected in Pt(4.0)/CoFeB(2.2) nanofilms no matter the pumping laser pulses are incident onto the samples in the case of $n^+$ and $n^-$ (see Figure 2c and d). The enhancement ratios are $P+_{Annealing}/P=2.3$ and $P-_{Annealing}/P=2.1$ for $n^+$ and $n^-$, respectively. The slightly differences between $n^+$ and $n^-$ in both W and Pt samples may be due to the various absorption of the glass substrate in optical frequency range and THz frequencies. If we ignore this small influence, we can safely conclude that the THz emission intensities of W(2.2)/CoFeB(2.0) and Pt(4.0)/CoFeB(2.2) nanofilms are enhanced by ~3 times and ~2 times by annealing effect, respectively, which are much higher than the optimized results reported in the previous work using a Ta(5.0)/CoFeB(1.0)/MgO(2.0) heterostructure.

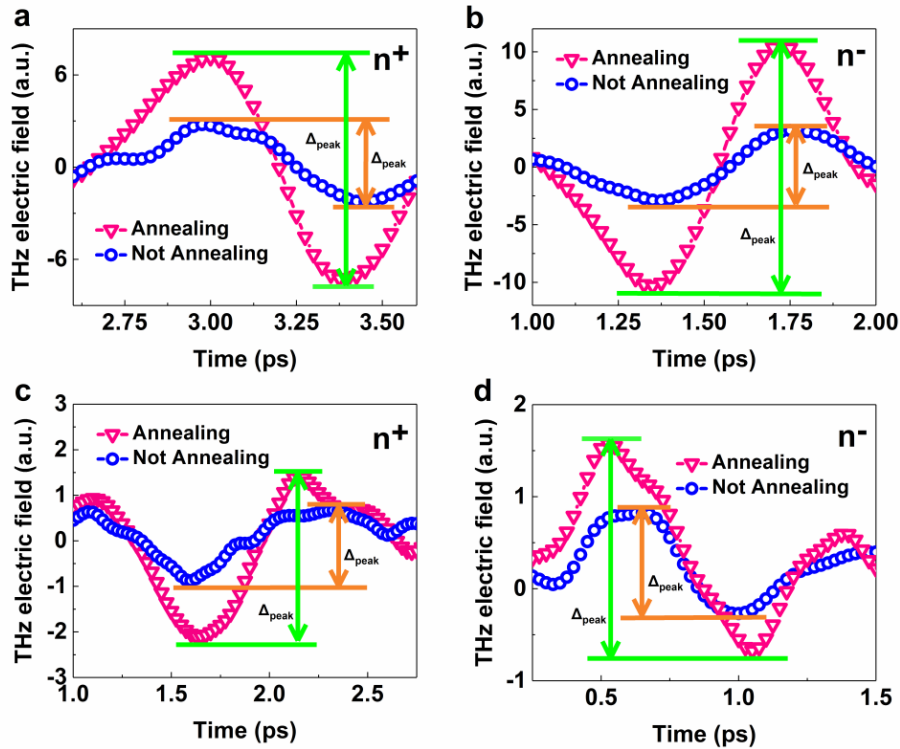

**Figure 2.** Enhanced THz emission intensity from bilayer samples after annealing. THz temporal waveforms from the W(2.2)/CoFeB(2.0) samples with and without annealing, for the cases of a, $n^+$ and b, $n^-$, respectively. c and d, Improved THz emission intensity are also observed in the Pt(4.0)/CoFeB(2.2) samples after annealing no matter the femtosecond laser pulses are incident from positive ($n^+$) or negative ($n^-$) sides.

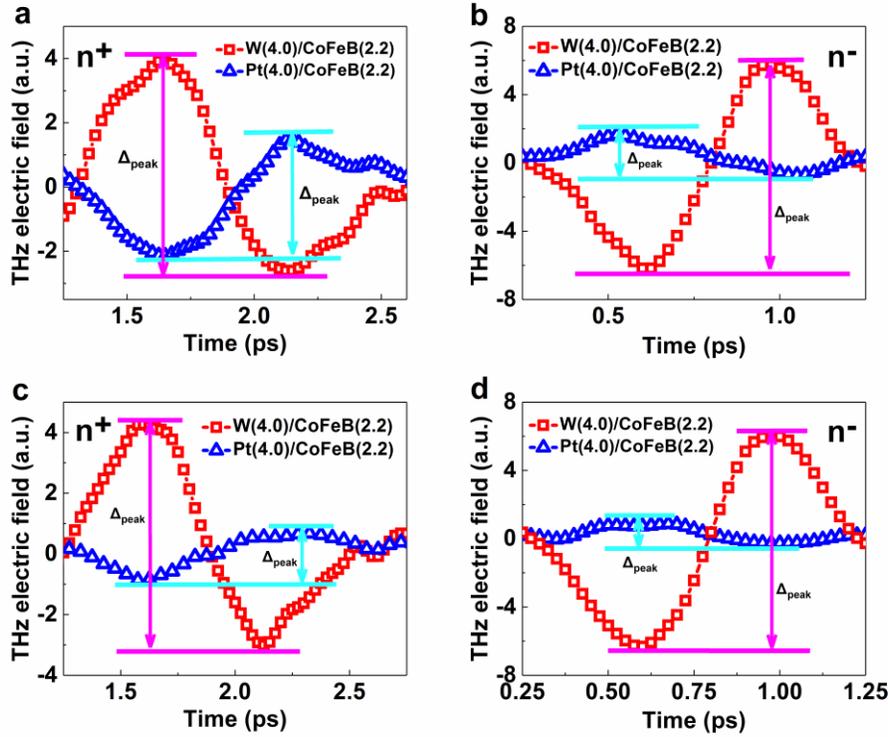

**Figure 3.** Eliminating the disturbance of the thickness dependent on THz emission intensity for verifying the annealing effect assisted THz enhancement. (a) and (b), Femtosecond laser pulses are incident onto the W(4.0)/CoFeB(2.2) and Pt(4.0)/CoFeB(2.2) nanofilms from the positive (n+) and negative (n-) sides, respectively, when both samples are annealed, while (c) and (d) illustrate the results from the both samples without annealing.

To eliminate the disturbance caused by the sample thickness, we select W(4.0)/CoFeB(2.2) and Pt(4.0)/CoFeB(2.2) films which are of the same thickness for the subsequent experiments. Figure 3a depicts the emitted THz signals from these two samples after annealing. From this figure, it can be seen that the THz emission phases are opposite which is attributed to the different polar of the spin Hall angles for W (negative) and Pt (positive) [7]. Additionally, the signal from W-based sample is ~1.9 time higher that that from Pt-based. This phenomenon may be due to the larger spin Hall angle of W than Pt [25]. This observed enhancement phenomenon is 5 times when the femtosecond laser pulses illuminate on the n- side. Likewise, in Figure 3c and 3d, the enhancement behavior is also observed and the ratio of peak-to-peak values of W(4.0)/CoFeB(2.2) and Pt(4.0)/CoFeB(2.2) nanofilms without annealing are $P_{W(4.0)}/P_{Pt(4.0)}$=4.76 and $P_{W(4.0)}/P_{Pt(4.0)}$=10.89, respectively. It is clearly manifested that the THz emission intensity of W-based emitters is always larger than the Pt-based ones, no matter they are annealed or not. Through these experiments, we can get rid of the thickness disturbance, and demonstrate that W is more appropriate than Pt as the HM layer of ferromagnetic heterostructures, which can bring greater THz emission intensity. Previously, *Seifert et al.* have already reported that W or Pt/CoFeB systems have the ability to achieve high emission efficiency due to their innate large spin Hall angles [7]. We achieve three times enhancement of THz emission intensity by extending the annealing effect to these efficient bilayer structures, according to the aforementioned results.

To gain a deep insight into the physics of annealing effect on THz emission enhancement, we will subsequently implement a comprehensive analysis on the possible mechanisms. Previously, one of the mechanisms was attributed to MgO contributed crystallization and enhancement of the spin Hall angle of Ta. In our case, as is shown in the sample preparation section, there is no MgO layer in our samples, but an intensity enhancement behavior is still observed. Therefore, this phenomenon may not be strongly correlated to this explanation. Then taking consideration into the other possibility, the crystallization of the ferromagnetic material layer leads to a reduction in atomic scale disorder caused by amorphous state, finally reducing spin-lattice scattering [26]. This effect can give rise to the increase of the mean free path for hot electrons, which has been verified annealing dependent THz emission experiments. Electrons carrying the majority spin state contribute more to the spin polarized current, the abovementioned change in the mean free path for hot electrons of majority spin will cause a reduction of spin current loss., thereby resulting in higher emission intensity. In order to corroborate this possible mechanism, verification experiments of THz transmission measurement on W(4.0)/CoFeB(2.2) and Pt(4.0)/CoFeB(2.2) with and without annealing are conducted in the THz time-domain spectrometer. As is shown in Figure 4a, it is obvious that the THz transmission intensity of W(4.0)/CoFeB(2.2) sample with annealing effect shows a slight enhancement compared with that without annealing. The intensity of transmitted THz signal has increased by 1.36%. Similar behavior in Pt(4.0)/CoFeB(2.2) is increased by 3.75%. These results corroborate the aforementioned second speculation. The increased transmitted THz wave may result from the decrease of energy dissipation which is caused by the scattering processes within a spin current induced by the incident THz wave. The decrease in spin-lattice scattering would have positive influence on this energy dissipation, namely, reservation of more energy. The crystallization of amorphous CoFeB caused by annealing effect reduces its lattice disorder, and then, depresses the spin-lattice scattering. Consequently, we can conclude that from this THz time-domain spectroscopy experiments, the improvement in THz transmission is a reasonable evidence for the decrease in energy-dissipating scattering process and increase in lattice order, thus may support the assumption that annealing effect promote the mean free path for hot electron, leading to a higher emission intensity. Although we detect a weak increase in THz transmission intensity for 1.36% in W(4.0)/CoFeB(2.2) and 3.75% in Pt(4.0)/CoFeB(2.2), respectively, which may due to the extremely thin thickness of CoFeB(2.2), the highest emission enhancement can be scaled up to 320%, demonstrating the considerably positive contribution of annealing to THz emission. Besides, we also conduct emission experiments on W(2.0)/CoFeB(2.2)/Pt(2.0) trilayer samples with and without annealing. However, we do not observe an obvious change of THz emission signals. We speculate that annealing parameters used in the experiments of W or Pt/CoFeB bilayers may not be suitable for W/CoFeB/Pt trilayer ferromagnetic films. The trilayer system is more complex than bilayers. As is illustrated in Figure 1b, c and e, bilayer samples are grown directly on the glass, but another HM layer is placed between FM layer and glass in this trilayer system. This may introduce extra interaction of annealing effects among FM layer, HM layer and glass substrate. Therefore, the proper annealing conditions required for trilayer samples may be different from the bilayers, but the annealing condition do not change in our sample preparation. To further clarify this issue, systematic studies on annealing conditions for trilayer samples need to be performed, which is beyond the scope of this work.

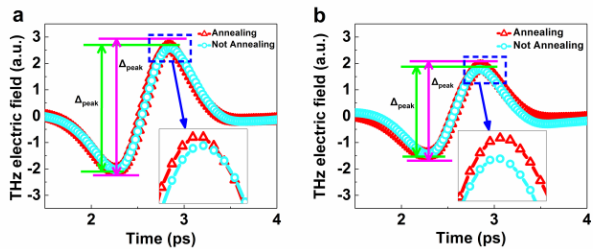

**Figure 4.** Increased THz transmission observed from the annealed samples. Transmitted THz time-domain signals in (a) W(4.0)/CoFeB(2.2) and (b) Pt(4.0)/CoFeB(2.2), with and without annealing. Inset: enlarged peak values.

## 3. CONCLUSION

In summary, we observe ~3 times enhanced THz emission intensity in W/CoFeB and Pt/CoFeB bilayers driven by femtosecond laser pulses. We narrowing the enhancement physics to annealing induced lattice disorder reduction which is reasonably supported by the THz time-domain transmission measurements. Through these experimental observations and qualitative analyses, we extend the annealing effect for demonstrated optimal spintronic THz sources. We believe our investigations not only help further deeply understand the fundamental physics of femtosecond spin dynamics, but also provide an optimization procedure for the highly efficient THz sources.

**Funding Information.** This work is supported by the National Natural Science Foundation of China (Grants No. 11827807, 11520101003, 11861121001, 61831001 and 61775233), the Strategic Priority Research Program of the Chinese Academy of Sciences (Grant No. XDB16010200 and XDB07030300), the National Basic Research Program of China (Grant No. 2014CB339800), the International Collaboration Project (Grant No. B16001), and the National Key Technology Program of China (Grant No. 2017ZX01032101). Dr. Xiaojun Wu thanks the "Zhuoyue" Program and "Qingba" Program of Beihang University (Grant No. GZ216S1807, ZG226S1832, KG12052501). Dr. Tianxiao Nie thanks the support from the 1000-Young talent program of China.